# Reliability Evaluation Method for Electronic Device BGA Package Considering the Interaction Between Design Factors

*Satoshi KONDO*, *Qiang YU*, *Tadahiro SHIBUTANI*,  
*Masaki SHIRATORI**

*Department of Mechanical Engineering and Materials Science  
Yokohama National University  
79-5, Tokiwadai, Hodogaya-ku, 240-8501, Japan  
Phone: +81-45-339-3862  
Fax: +81-45-331-6593  
Email: qiang@swan.me.ynu.ac.jp

**ABSTRACT**

The recent development of electric and electronic devices has been remarkable. The miniaturization of electronic devices and high integration are progressing by advances in mounting technology. As a result, the reliability of fatigue life has been prioritized as an important concern, since the thermal expansion difference between a package and printed circuit board causes thermal fatigue. It is demanded a long-life product which has short development time. However, it is difficult because of interaction between each design factor. The authors have investigated the influence of various design factors on the reliability of soldered joints in BGA model by using response surface method and cluster analysis. By using these techniques, the interaction of all design factors was clarified. Based upon the analytical results, design engineers can rate each factor's effect on reliability and assess the reliability of their basic design plan at the concept design stage.

## 1. INTRODUCTION

Recently, in development of electronic devices, shortening of a development period and reduction of cost becomes more important subject. On the other hand, at the same time, a guarantee of quality that user expects should be obtained. From these demands, an efficient design support system for electronics devices is expected Advancing of mounting technology, electronic devices were miniaturized and integrated. Because of this, solder joints of devices were detailed, and heat fatigue destruction of solder joint became a serious problem. This problem was caused by the difference of thermal expansion between a package and printed circuit board. And it is necessary to try to improve this problem at the design stage. However, because of complicated structure of electronic devices, the interaction of each design factor became remarkable and reliability problem has been complicated. So, it is very difficult to give effective changes with design factors for heat fatigue life. Therefore, the design support tool corresponding to the complicated reliability problem is needed.

The purpose of this study is to establish the simple and convenience design support technique for electronics devices. As its application, the influence and interaction of each design factor in BGA package on thermal fatigue life was examined. Since BGA package has complicated structure. To investigate the influence of design factors on heat fatigue life, sensitivity analysis was used. But in cases that the package has complicated structure, it is known that sensitivity analysis becomes not much suitable because of the interaction between each design factor. So, in order to clarify this interaction, cluster analysis was used. Furthermore, the technique for clarifying a more detailed interaction of each design factor was also examined. As a result, all interaction was clarified and more collect sensitivity analysis was done.

The contents of this study are stated as follows:

1)   Influence analysis by using surface response method
2)   Cluster analysis for clarifying the interaction of each design factor
3)   Cluster analysis, which observe one factor and clarifies more detailed interaction

By using these methods, interaction of each design factor is clarified. Moreover, exact sensitivity analysis can be performed by taking interaction into consideration.

## 2. NOMENCLATURE

   FEM      Finite Element Method
   BGA      Ball Grid Array
   PCB      Printed Circuit Board

CTE Coefficient of Thermal Expansion
DOE Design Of Experiment

## 3. THE INFLUENCE ANALYSIS ON DESIGN FACTOR TO TOTAL EQUIVALENT INELASTIC STRAIN RANGE

Several methods can be used to evaluate the thermal fatigue life of solder joints [1-2]. In this study, the initial fatigue crack occurring in the solder joints is used for the thermal fatigue life [3-4]. It is known that the total equivalent inelastic strain range per step of thermal load can be used to evaluate the life [5-8].

To clarify the relation between design factors and total equivalent inelastic strain range, sensitivity analysis was used. At first, in the model of fundamental BGA package as shown in Fig. 1, sizes of components, mechanical properties and thermal properties were taken as the design factor shown in Table. 1. Orthogonal table was created by the DOE theory. In this case study, total equivalent inelastic strain range was taken as the characteristic value calculated by using FEM analysis. The detail condition of FEM analysis is stated as follows:

1) The temperature range is from -40ºC to +125ºC for the thermal load. The time of temperature change take 0.05 hour (3 minutes), and the dwelling time is 0.25 hour (15 minutes) as shown in Fig. 2.
2) Based upon the symmetry of the package structure, a quarter model was used in this analysis, and the symmetrical boundary conditions are subjected as show in Fig.3.
3) The total equivalent inelastic strain range was calculated with the average equivalent strains around 50μm at the corner of the solder bump shown in the circle of Fig. 4.

By performing response surface method, a characteristic value can be expressed on the basis of estimated equation as shown in Table.2. And the influence of the design factor to a total equivalent inelastic strain range was calculated. The table of influence for design factors in Table.3. From this result, it makes clear that the thickness of Encap had affected the most, and in order of influence, CTE and thickness of substrate and Encap's CTE also affect characteristic value. Moreover, the influence figure of a design factor was shown in Fig. 5. We can identify easily how much influence each design factor has on a characteristic value from this figure. This shows that thickness of Encap, CTE of PCB, and CTE of Encap should be made into a level 1 (low value), and thickness of substrate, Young's modulus of substrate, and CTE of substrate must be made into a level 3 (high value) in order to make a characteristic value low.

By these studies, the relation between design factors and a heat fatigue life becomes clearer. However, this technique is not taken into consideration about the interaction of each design factors. To clarify the interaction of each design factor, clustering analysis shown below was used.

## 4. CLUSTER ANALYSIS

To clarify the interaction of each design factor, cluster analysis was used in this study. Cluster Analysis is the method of calculating the Euclid distance between parameters, and gathering close models one after another, and expressing their relation by a hierarchical structure. Clustering similar results to the characteristic value acquired in FEM analysis etc., and the interaction of each design factor to a characteristic value makes it clear for comparing each cluster.

The values of total equivalent inelastic strain range obtained from FEM analysis based on the orthogonal table were plotted in Fig. 6. A vertical axis shows the total equivalent inelastic strain range that which is a characteristic value, and the horizontal axis shows the data number. Those data were arranged in order of total equivalent inelastic strain range, and the close models have been clustered to four clusters by Euclid distance. In order to investigate the relation of each design factor, each design factor was averaged in a cluster and these values of design factors in each cluster are plotted in Fig. 7. The X-axis shows design factor and the Y-axis shows the design value regularized by design range. No.1 cluster in figure shows the data of a design factor pattern when the value of total equivalent inelastic strain range is smaller, and No.4 cluster shows the data of a design factor pattern when total equivalent inelastic strain range is bigger. The arrow shows the direction where total equivalent inelastic strain range becomes large, and its length indicates the intensity of influence to the characteristic value. From this result, the changing of design factor when total equivalent inelastic strain range was small or big was clarified, and it is possible to grasp all design factor differences at once. Therefore, the whole interaction can be grasped by using this figure.

From the result of this cluster, as the thickness and CTE of Encap become smaller, and the thickness and CTE of substrate are made larger, the total equivalent inelastic strain range tend to decrease. In the case of this condition, the trend of Encap's properties shows the opposite to substrate's one. In previous paper, it was reported that whole package would curve upwards [9]. And the total equivalent inelastic strain range of a solder joint tends to decrease when packages curve upwards as shown in Fig.

8. Furthermore, in the case of thickness of chip factor, when the total equivalent inelastic strain range is high, or low, the value of chip height is both large. This means that the height of a chip is influenced considerably with the other design factor. From this result, it is clear that the influence analysis is not taking interaction of each design factor into consideration and obtained result is a primary. This clustering clarifies the whole interaction of each design factor, but it is still not clear which factors have a interaction concretely. In order to perform a more exact development, it is important to take the detail interaction of design factors into consideration. Then, paying attention to one factor, the method to do clarification with more detailed interaction of each design factor would be examined.

## 5. DETAIL ANALYSIS OF INTERACTION BETWEEN DESIGN FACTORS

When there is change added to one certain factor, the method of investigating the influence of other factors was examined. At first, from the result of Fig. 5, cases where the value of thickness of substrate takes maximum or minimum were extracted. And it was arranged in order of the size of total equivalent inelastic strain range as shown in Fig. 9. The X-axis and the Y-axis takes the same as Fig. 5, respectively. The points that are painted show the results of minimum value of thickness of substrate, and the points that are not painted show the result of maximum value of thickness of substrate. By clustering similar results to the characteristic value and comparing the average of each design factor in each cluster, the effect of changing of thickness of substrate was clarified.

The result is shown in Fig. 10. This figure means the influence of changing thickness of substrate value on another design factor. When thickness of substrate is low, thickness of chip should be made low in order to decrease the total equivalent inelastic strain range. But when thickness of substrate is high, it should be higher. It turned out that, when one design factor (in this case, substrate's height) was changed, the influence of another design factor (chip's height) on the characteristic value was reversed. In similar case, when thickness of substrate is low, the factor of CTE of substrate doesn't influence the characteristic value. But when thickness of substrate is high, it influences the characteristic value. Therefore, it became clearer that thickness of chip and CTE of substrate have strong interaction with thickness of substrate. And interaction strength can be calculated from the difference between regularized values (arrows in the figure).

To apply this clustering technique to all design factors, all interaction of each design factor would be clarified.

## 6. EXAMPLE OF APPLYING THIS CLUSTERING TECHNIQUE

An example of applying this technique is shown below. From the previous study, it is known that the reliability assessment could be done easily by calculating the appearance CTE and Young's modulus of whole BGA package [9]. To estimate the appearance CTE of whole BGA package, the interaction between all design factors was calculated and sensitivity analysis was done. Table.4 shows the interaction strength between all design factors calculated by the clustering method. The strength of the interaction relation between every two factors can be expressed by a sum of the correlation coefficient at the cross points of the two factors. When the sum value is bigger then the average of the all sum values, the two factors have interaction relation, and considering this result, sensitivity analysis was evaluated and estimated equation was calculated. At the same time, sensitivity analysis which considers no interaction was executed, and both estimated equations are compared as shown in Table.5.

Two different FEM models of various design conditions were made, and appearance CTE of whole package was calculated from the FEM analysis, estimated equation with considering interaction, and estimated equation without considering interaction. By comparing with the result of FEM analysis, it is clarified that estimated equation with interaction is more accurate than the other one. From this result, effectiveness of this technique was clarified.

## 7. CONCLUSION

1) By using the influence analysis, the influence of each design factor on thermal fatigue life in electronics devices can be evaluated quantitatively. However, because of interaction of each design factor, this method is not much suitable.
2) Clustering technique can provide the whole interaction of each design factor.
3) Moreover, by clustering for the observed design factor, detail interaction of each design factor can be extracted.
4) It was confirmed that exact sensitivity analysis was performed by considering the interaction.

As a result, it was proved that all interaction of each design factor could be clarified and more exact sensitivity analysis could be performed. Therefore, design engineers can rate each factor's effect on reliability and assess the reliability of their basic design plan at the concept design stage.

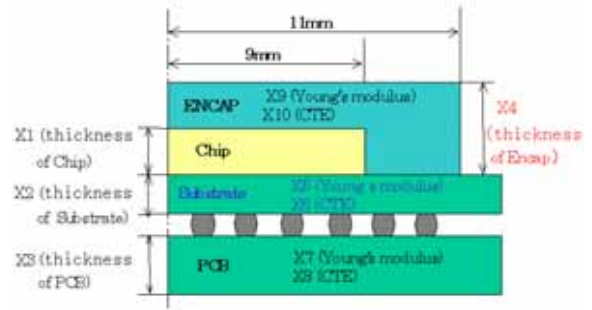

Fig. 1 Package model

Table 1. Package design factors and value levels

| Factor | Min | Ave | Max |
|---|---|---|---|
| Thickness of Chip(μm) | 300 | 400 | 500 |
| Thickness of Sub'(μm) | 300 | 400 | 500 |
| Thickness of PCB(μm) | 800 | 1000 | 1200 |
| Thickness of Encap(μm) | 1000 | 1200 | 1400 |
| Young's modulus of Sub'(GPa) | 15 | 19 | 23 |
| CTE of Sub'(ppm/K) | 12 | 15 | 18 |
| Young's modulus of PCB(GPa) | 15 | 19 | 23 |
| CTE of PCB(ppm/K) | 13 | 16 | 19 |
| Young's modulus of Encap(GPa) | 13 | 16 | 19 |
| CTE of Encap(ppm/K) | 12 | 15 | 18 |

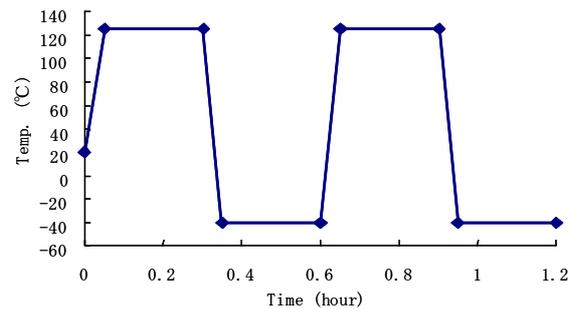

Fig. 2. Thermal load of analysis

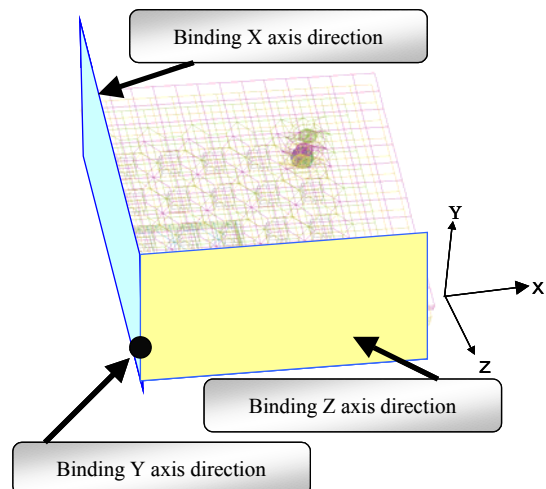

Fig. 3. Boundary conditions

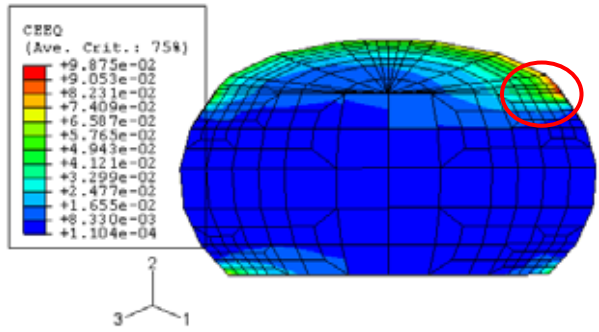

Fig.4 FEM model of solder bump

Table.2 Estimated equation

$\Delta \varepsilon$ in = $-5.97*10^{-3}+4.36*10^{-5}*X1-4.91*X1^2-1.32*10^{-5}*X2-7.79*10^{-9}*X2^2+3.63*10^{-5}*X4-6.94*10^{-9}*X4^2+3.40*X7-1.01*10^{-5}*X7^2+2.06*10^{-5}*X3-1.03*10^{-8}*X3^2-4.98*10^{-4}*X5+3.47*10^{-7}*X5^2-1.33*10^{-3}*X8+5.99*X8^2+6.02*10^{-4}*X6-5.12*10^{-5}*X6^2-7.20*10^{-4}*X9+3.21*10^{-5}*X9^2-9.78*10^{-4}*X10+5.43*10^{-5}*X10^2$

*X1~X10: design factors

Table 3. Influence of the package model

| Factor | Effective ratio |
|---|---|
| Thickness of Chip | 0.39% |
| Thickness of Sub' | 9.43% |
| Thickness of PCB | 0.03% |
| **Thickness of Encap** | **38.90%** |
| Young's modulus of Sub' | 9.37% |
| **CTE of Sub'** | **19.81%** |
| Young's modulus of PCB | 0.00% |
| CTE of PCB | 7.94% |
| Young's modulus of Encap | 2.03% |
| CTE of Encap | 9.63% |
| Error | 2.44% |
| Total | 100.00% |

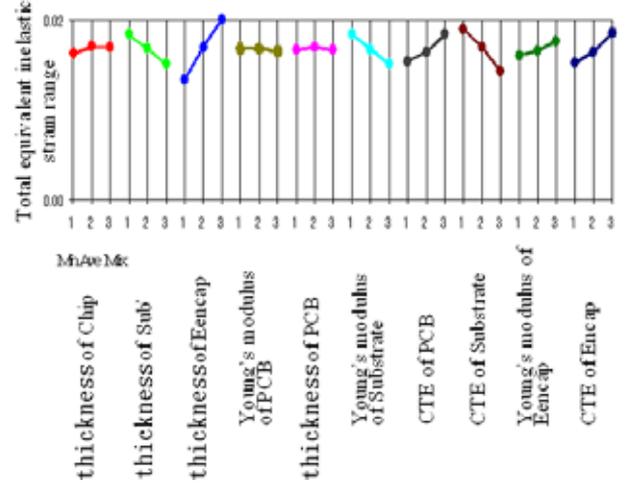

Fig. 5. Influence figure of a design factor

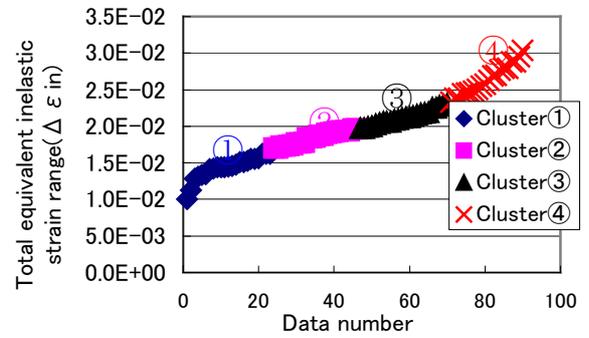

Fig. 6. Data of total equivalent inelastic strain range

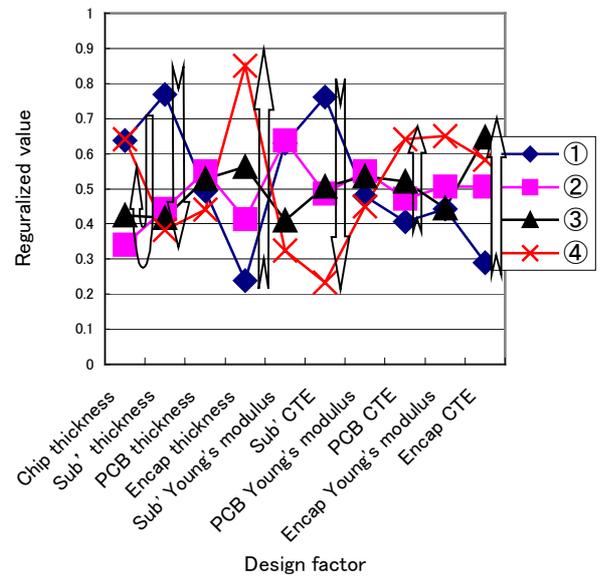

Fig.7. Clustering of BGA package design factor

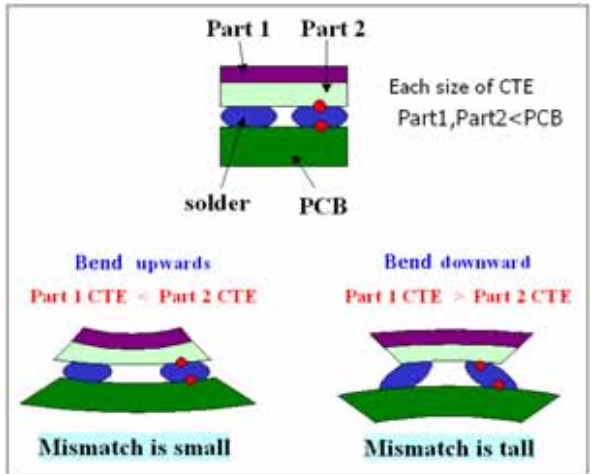

Fig.8. Structure of influence in the solder by curvature

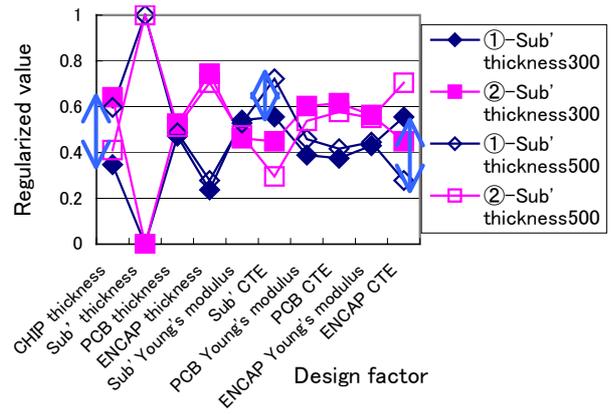

Fig. 10. Clustering when substrate thickness changed

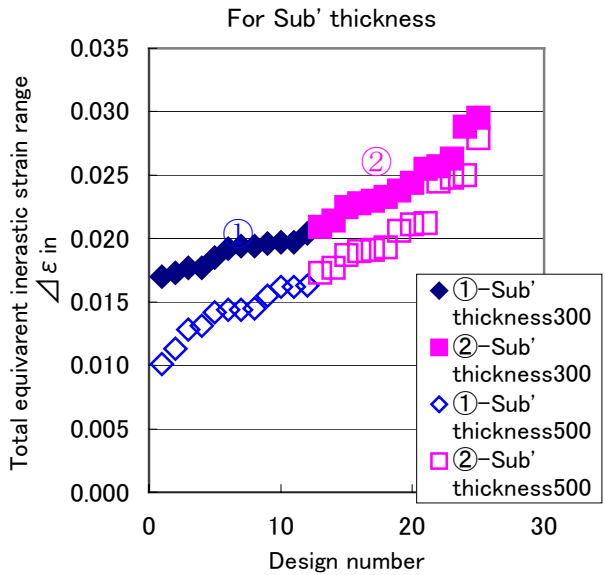

Fig. 9.Data of total equivalent inelastic strain range in case
substrate thickness is maximum and minimum

Table.4 All correlation coefficient of BGA package

| | Encap CTE | Encap Young's | Encap thickness | Chip CTE | Chip Young's | Chip thickness | Sub' CTE | Sub' Young's | Sub' thickness |
|---|---|---|---|---|---|---|---|---|---|
| Encap CTE | - | 0.04 | 0.00 | 0.10 | 0.02 | 0.55 | 0.06 | 0.33 | -0.05 |
| Encap Young's | 0.12 | - | -0.29 | -0.20 | 0.04 | 0.04 | -0.02 | 0.02 | 0.03 |
| Encap thickness | 0.10 | -0.23 | - | 0.06 | -0.16 | 0.15 | 0.04 | 0.05 | 0.12 |
| Chip CTE | 0.11 | -0.16 | 0.19 | - | -0.11 | 0.04 | -0.07 | -0.06 | 0.01 |
| Chip Young's | 0.06 | 0.04 | -0.15 | -0.22 | - | 0.02 | -0.03 | -0.03 | 0.06 |
| Chip thickness | 0.32 | -0.13 | 0.08 | 0.02 | 0.00 | - | 0.06 | -0.04 | -0.02 |
| Substrate CTE | 0.05 | 0.05 | 0.04 | 0.04 | 0.02 | -0.10 | - | 0.10 | -0.18 |
| Substrate Young's | 0.19 | 0.06 | -0.03 | -0.05 | 0.02 | 0.10 | 0.04 | - | -0.04 |
| Substrate thickness | 0.09 | 0.00 | 0.26 | 0.10 | 0.04 | -0.05 | -0.19 | -0.04 | - |

Table.5 Results from estimated equations

| | Design factors | Min. value | Max. value | Model 1 | Model 2 |
|---|---|---|---|---|---|
| X1 | CTE of encap ($10^{-6}$/°C) | 12 | 20 | 15 | 16 |
| X2 | Young's modulus of encap (GPa) | 15 | 23 | 18 | 15.5 |
| X3 | Thickness of encap (mm) | 0.2 | 0.6 | 0.58 | 0.23 |
| X4 | CTE of chip ($10^{-6}$/°C) | 2 | 6 | 3 | 6 |
| X5 | Young's modulus of chip (GPa) | 150 | 200 | 180 | 175 |
| X6 | Thickness of chip (mm) | 0.1 | 0.5 | 0.44 | 0.29 |
| X7 | CTE of substrate ($10^{-6}$/°C) | 12 | 20 | 16 | 17 |
| X8 | Young's modulus of substrate (GPa) | 15 | 23 | 16 | 22.5 |
| X9 | Thickness of substrate (mm) | 0.1 | 0.5 | 0.48 | 0.47 |
| | Result | | | CTE ($10^{-6}$/°C) | CTE ($10^{-6}$/°C) |
| | FEM analysis | | | 4.33 | 15.8 |
| | Not considering interaction | | | 7.15 | 13.9 |
| | Considering interaction | | | 4.11 | 15.6 |